\documentclass[twocolumn,final,10pt]{IEEEtran}
\usepackage{graphicx}

\usepackage{cite}
\usepackage{amsmath}
\usepackage{amsthm}
\usepackage{subfigure}
\usepackage{float}
\usepackage{times}
\usepackage{enumerate}
\usepackage{amssymb}
\usepackage{algorithm}
\usepackage{algorithmic}
\usepackage{multirow}
\usepackage{braket}
\usepackage{multicol}
\usepackage{caption}
\newtheorem{theorem}{Theorem}
\newtheorem{definition}{Definition}

\begin{document}

\title{On the Stability Region of a Relay-Assisted Multiple Access Scheme}

\author{\authorblockN{Nikolaos Pappas\textsuperscript{*}, Marios Kountouris\textsuperscript{*}, Anthony Ephremides\textsuperscript{\ddag}, Apostolos Traganitis\textsuperscript{\dag}}\\
\authorblockA{\textsuperscript{*} Sup\'{e}lec, Department of Telecommunications, Gif-sur-Yvette, France\\
\textsuperscript{\ddag} Department of Electrical and Computer Engineering and Institute for Systems Research\\
University of Maryland, College Park, MD 20742\\
\textsuperscript{\dag}Computer Science Department, University of Crete, Greece\\Institute of Computer Science, Foundation for Research and Technology - Hellas (FORTH)\\
Email: \{nikolaos.pappas, marios.kountouris\}@supelec.fr, etony@umd.edu, tragani@ics.forth.gr
}

\thanks{This research has been partly supported by the ERC Starting Grant 305123 MORE (Advanced Mathematical Tools for Complex Network Engineering).}
\thanks{This work has been supported in part by the EU-FP7 Marie Curie IAPP project MESH-WISE.}}

\maketitle

\begin{abstract}
In this paper we study the impact of a relay node in a two-user network. We assume a random access collision channel model with erasures. In particular we obtain an inner and an outer bound for the stability region.
\end{abstract}

\section{Introduction} \label{sec:intro}
The classical relay channel was originally introduced by van der Meulen~\cite{b:Muelen}. Earlier work on the relay channel was based on information theoretical formulations such as in~\cite{b:CoverGamal}. Most cooperative techniques studied so far have been on physical layer cooperation that enables non-trivial benefits~\cite{b:Yates-NOW}. Nevertheless, there is evidence that the same gains can be achieved with network layer cooperation, which is plain relaying without any physical layer considerations~\cite{b:Sadek, b:Rong1}.

Recently several works have investigated relaying performance at the medium access control layer~\cite{b:Sadek, b:Rong1, b:Rong2}. More specifically, in~\cite{b:Sadek}, the authors have studied the impact of cooperative communication at the medium access control layer with TDMA. They introduced a new cognitive multiple access protocol in the presence of a relay in the network. The authors in~\cite{b:Pappas-ISIT} introduced the notion of partial network level cooperation by adding a flow controller at the relay, which regulates the amount of provided cooperation depending on the conditions of the network.

A key difference between physical layer and network layer cooperation is that in the second case the objective rate function that is maximized is the so-called stable throughput region (also called stability region), which captures the bursty nature of traffic.

In this work, we study a network which consists of two user sources with external arrivals and one relay, where the relay does not have packets on its own. The relay is forwarding part of the sources' traffic, hence the arrival rate at the relay is dependent on the arrival rates at the sources. We consider a collision channel with erasures and random medium access. The characterization of the stability region of random access systems is known to be a very challenging problem. The difficulty lies on the interaction (or coupling) among the queues i.e., the service process of a queue depends on the status of the other queues. This is the reason why most of the studies focus on small-sized networks~\cite{rao:stability, Szpankowski:stability} and only bounds are known for networks with large number of nodes~\cite{luo:stability}.
For the characterization of the stability region, we obtain an inner and an outer bound based on the stochastic dominance technique for decoupling the queues~\cite{rao:stability}.

This paper is organized as follows. The system model is described in Section~\ref{sec:model}. The main results are given in Section~\ref{sec:theorems}, and their proofs are given in Sections~\ref{sec:analysis_inner} and~\ref{sec:analysis_outer}. Numerical results are presented in Section~\ref{sec:results} and the paper is concluded in Section~\ref{sec:conclusions}.

\section{System Model} \label{sec:model}
\begin{figure}[t]
\centering
\includegraphics[scale=1.1]{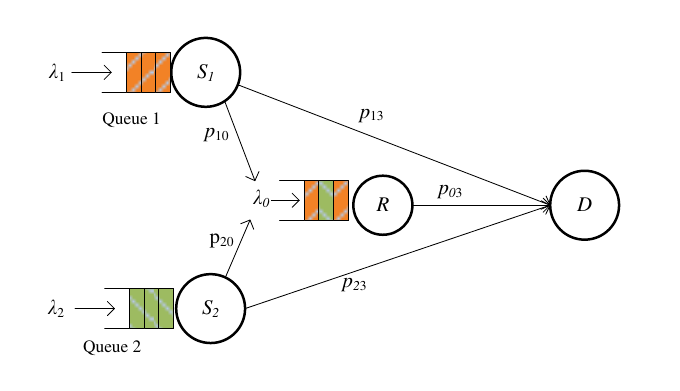}
\caption{Relay-Assisted Multiple Access Channel}

\label{fig:model}
\end{figure}

We consider a time-slotted system in which the nodes are randomly sending to a common receiver as shown in Fig.~\ref{fig:model}. We denote with $S_i$, $R$ and $D$, the source-user $i$ ($i=1,2$), the relay and the destination, respectively. Traffic is originated only from $S_1$ and $S_2$. The relay node is a pure relay and does not generate traffic on its own. Due to the broadcast nature of the wireless medium, $R$ may receive and decode some of the packets transmitted from the sources $S_i$ so as to relay them to $D$.

Packets from $S_i$, which failed to be received by $D$ but were successfully received by $R$, are relayed by $R$. As half-duplex communication is assumed here, $R$ can overhear $S$ only when it is idle. Each node has an infinite size buffer for storing incoming packets and the transmission of each packet occupies one time slot.

The packet arrival processes at $S_1$ and $S_2$ are assumed to be Bernoulli with rates $\lambda_1$ and $\lambda_2$, respectively, and are independent of each other. The arrival process $\lambda_0$ at the relay, which is a linear combination of $\lambda_1$ and $\lambda_2$, is shown in the next section.

In each time slot, nodes $S_1$, $S_2$ and $R$ attempt to transmit with probabilities $q_1$, $q_2$ and $q_0$, respectively, if their queues are not empty. Decisions on transmission are made independently among nodes. We assume a collision channel with erasures in which, if more than one node (source or relay) transmits in the same time slot, a collision occurs and all transmissions fail. The probability that a packet transmitted by node $i$ is successfully decoded at node $j (\neq i)$ is denoted by $p_{ij}$ which is the probability that the signal-to-noise-ratio (SNR) over the specified link exceeds a certain threshold for successful decoding. These erasure probabilities capture the effect of random fading at the physical layer. The probabilities $p_{i3}$, $p_{03}$, and $p_{i0}$ denote the success probabilities over the links $S_i-D$, $R-D$, and $S_i-R$, respectively. Node $R$ has a better channel to $D$ than $S_i$, that is $p_{03} > p_{i3}$, otherwise the presence of the relay will degrade the performance of the whole network.

The cooperation is performed at the protocol level as follows. When $S_i$ transmits a packet, if $D$ decodes the packet successfully, it sends an ACK and the packet exits the network; if $D$ fails to decode the packet but $R$ does, then $R$ sends an ACK and takes over the responsibility of delivering the packet to $D$ by placing it in its queue. If neither $D$ nor $R$ decode (or if $R$ does not store the packet), the packet remains in $S_i$'s queue for retransmission. The ACKs are assumed to be error-free, instantaneous and broadcasted to all relevant nodes.

We use the following definition of queue stability~\cite{Szpankowski:stability}:

\begin{definition}
Denote by $Q_i^t$ the length of queue $i$ at the beginning of time slot $t$. The queue is said to be \emph{stable} if
\begin{equation*}\label{eqn:definition_stability}
    \lim_{t \rightarrow \infty} {Pr}[Q_i^t < {x}] = F(x)  \text{ and } \lim_{ {x} \rightarrow \infty} F(x) = 1.
\end{equation*}

If

\begin{equation*}
     \lim_{x \rightarrow \infty}  \lim_{t \rightarrow \infty} \inf {Pr}[Q_i^t < {x}] = 1
\end{equation*}

the queue is \emph{substable}. If a queue is stable, then it is also substable. If a queue is not substable, then we say it is unstable.
\end{definition}

Loynes' theorem~\cite{b:Loynes} states that if the arrival and service processes of a queue are strictly jointly stationary and the average arrival rate is less than the average service rate, then the queue is stable. If the average arrival rate is greater than the average service rate, then the queue is unstable and the value of $Q_i^t$ approaches infinity almost surely. The stability region of the system is defined as the set of arrival rate vectors $\boldsymbol{\lambda}=(\lambda_1, \lambda_2)$ for which the queues in the system are stable.

Note that in this problem we have three queues, however we have only two external traffic sources, thus the stability region in our problem is two-dimensional (all of the three queues must be stable).

\section{Main Results} \label{sec:theorems}

This section states the main results of this paper, which are inner and outer bounds for the stability region of the system presented in the previous section (cf. Fig.~\ref{fig:model}).

\begin{theorem} \label{thm:th1}
An inner bound for the stability region of the relay-assisted multiple access network depicted in Fig.~\ref{fig:model} is given by
\begin{equation}
    \mathcal{R}_{IN} = \left( \mathcal{R}_{IN}^1 \bigcup \mathcal{R}_{IN}^2 \right)  \bigcap \mathcal{R}_{IN}^R
\end{equation}

where the subregions $\mathcal{R}_{IN}^1$ and $\mathcal{R}_{IN}^2$ are given by (\ref{eq:R_IN_1}) and (\ref{eq:R_IN_2}), respectively.
The subregion $\mathcal{R}_{IN}^R$, which is the condition for the relay to be stable, is given by (\ref{eq:R_IN_R}).

\begin{proof}
The proof is given in Section~\ref{sec:analysis_inner}.
\end{proof}
\end{theorem}

\begin{figure*}[!t]
\begin{equation} \label{eq:R_IN_1}
\begin{aligned}
\mathcal{R}_{IN}^1 = \left( (\lambda_{1},\lambda_{2}): \frac{(1-q_1)(1-q_2)p_{03}+q_1 p_{10}(1-p_{13})}{q_1(1-q_1)(1-q_2)p_{03}[p_{13}+p_{10}(1-p_{13})]} \lambda_1 +\frac{(1-q_2)p_{03}+p_{20}(1-p_{23})}{(1-q_1)(1-q_2)p_{03}[p_{23}+p_{20}(1-p_{23})]} \lambda_2 < 1, \right. \\
\left. \frac{p_{10}(1-p_{13})}{p_{13}+p_{10}(1-p_{13})} \lambda_1 + \frac{(1-q_2)p_{03}+q_2p_{20}(1-p_{23})}{q_2[p_{23}+p_{20}(1-p_{23})]} \lambda_2 < (1-q_1) (1-q_2) p_{03} \right)
\end{aligned}
\end{equation}
\begin{equation} \label{eq:R_IN_2}
\begin{aligned}
\mathcal{R}_{IN}^2= \left( (\lambda_{1},\lambda_{2}): \frac{(1-q_1)p_{03}+p_{10}(1-p_{13})}{(1-q_1)(1-q_2)p_{03}[p_{13}+p_{10}(1-p_{13})]} \lambda_1 + \frac{(1-q_1)(1-q_2)p_{03}+q_2 p_{20}(1-p_{23})}{q_2 (1-q_1)(1-q_2)p_{03}[p_{23}+p_{20}(1-p_{23})]} \lambda_2 < 1, \right. \\
\left. \frac{q_1 p_{10} (1-p_{13})+(1-q_1) p_{03}}{q_1 [p_{13}+p_{10}(1-p_{13})]} \lambda_1 + \frac{p_{20}(1-p_{23})}{p_{23}+p_{20}(1-p_{23})} \lambda_2 < (1-q_1) (1-q_2) p_{03}\right)
\end{aligned}
\end{equation}

\begin{equation} \label{eq:R_IN_R}
\begin{aligned}
\mathcal{R}_{IN}^R = \left( (\lambda_{1},\lambda_{2}): \frac{p_{10}(1-p_{13})}{p_{13}+p_{10}(1-p_{13})} \lambda_1 + \frac{p_{20}(1-p_{23})}{p_{23}+p_{20}(1-p_{23})} \lambda_2 < q_0 (1-q_1) (1-q_2) p_{03} \right)
\end{aligned}
\end{equation}

\begin{equation} \label{eq:R_OUT_1}
\mathcal{R}^{1}_{OUT} = \left( (\lambda_{1},\lambda_{2}):\frac{\lambda_1}{q_1[p_{13}+p_{10}(1-p_{13})]}+\frac{\lambda_2}{(1-q_1)[p_{23}+p_{20}(1-p_{23})]} < 1 ,\lambda_2 < q_2 (1-q_1)[p_{23}+p_{20}(1-p_{23})] \right)
\end{equation}
\begin{equation} \label{eq:R_OUT_2}
\mathcal{R}^{2}_{OUT} = \left(\lambda_{1},\lambda_{2}):\frac{\lambda_2}{q_2[p_{23}+p_{20}(1-p_{23})]}+\frac{\lambda_1}{(1-q_2)[p_{13}+p_{10}(1-p_{13})]} < 1 ,\lambda_1 < q_1 (1-q_2)[p_{13}+p_{10}(1-p_{13})] \right)
\end{equation}

\begin{equation} \label{eq:R_OUT_R}
\mathcal{R}_{OUT}^R = \left( (\lambda_{1},\lambda_{2}): \frac{p_{10}(1-p_{13})}{p_{13}+p_{10}(1-p_{13})} \lambda_1 + \frac{p_{20}(1-p_{23})}{p_{23}+p_{20}(1-p_{23})} \lambda_2 < p_{03} \right)
\end{equation}
\end{figure*}

\begin{theorem} \label{thm:th2}
An outer bound for the stability region of the relay-assisted multiple access network depicted in Fig.~\ref{fig:model} is given by
\begin{equation}
    \mathcal{R}_{OUT} = \left( \mathcal{R}_{OUT}^1 \bigcup \mathcal{R}_{OUT}^2 \right) \bigcap \mathcal{R}_{OUT}^R
\end{equation}

where the subregions $\mathcal{R}_{OUT}^1$ and $\mathcal{R}_{OUT}^2$ are given by (\ref{eq:R_OUT_1}) and (\ref{eq:R_OUT_2}), respectively.
The subregion $\mathcal{R}_{OUT}^R$, which is the condition for the relay to be stable, is given by (\ref{eq:R_OUT_R}).

\begin{proof}
The proof is given in Section~\ref{sec:analysis_outer}.
\end{proof}
\end{theorem}

The equations (\ref{eq:R_IN_1})-(\ref{eq:R_OUT_R}) are given on the top of the next page.

\section{Inner Bound For the Stability Region: Stochastic Dominance Analysis} \label{sec:analysis_inner}

In order to derive the stability conditions for the queues, we need to calculate the total arrival rate $\lambda_0$ in the relay node.
Let $\mathcal{A}_1$ ($\mathcal{A}_2$) denote the event that $S_1$ ($S_2$) transmits a packet and the packet leaves the queue. Then, we have that:

\begin{equation}
\mathrm{Pr}(\mathcal{A}_1)= (1-q_2\mathrm{Pr}[Q_2 > 0]) (1-q_0\mathrm{Pr}[Q_0 > 0]) [p_{13}+p_{10}(1-p_{13})]
\end{equation}

\begin{equation}
\mathrm{Pr}(\mathcal{A}_2)= (1-q_1\mathrm{Pr}[Q_1 > 0]) (1-q_0\mathrm{Pr}[Q_0 > 0]) [p_{23}+p_{20}(1-p_{23})].
\end{equation}

Among the packets that depart from the queue of $S_1$ ($S_2$), some will depart from the network since they are decoded by the destination directly, and
some will be relayed by $R$. Denote by $\mathcal{B}_1$ ($\mathcal{B}_2$) the event that the transmitted packet from $S_1$ ($S_2$) will be relayed by $R$, then we have:

\begin{equation}
\mathrm{Pr}(\mathcal{B}_1)= (1-q_2\mathrm{Pr}[Q_2 > 0]) (1-q_0\mathrm{Pr}[Q_0 > 0]) p_{10}(1-p_{13})
\end{equation}

\begin{equation}
\mathrm{Pr}(\mathcal{B}_2)= (1-q_1\mathrm{Pr}[Q_1 > 0]) (1-q_0\mathrm{Pr}[Q_0 > 0]) p_{20}(1-p_{23}).
\end{equation}

The conditional probabilities that a transmitted packet from $S_1$, $S_2$ arrives at $R$ given that the transmitted packet departs from nodes' $S_1$, $S_2$ queues are given by

\begin{equation}
\mathrm{Pr}(\mathcal{B}_1 | \mathcal{A}_1)=\frac{(1-p_{13})p_{10}} {p_{13}+(1-p_{13})p_{10}}
\end{equation}

\begin{equation}
\mathrm{Pr}(\mathcal{B}_2 | \mathcal{A}_2)=\frac{(1-p_{23})p_{20}} {p_{23}+(1-p_{23})p_{20}}.
\end{equation}

The total arrival rate at the relay is a linear combination of $\lambda_1$ and $\lambda_2$ and is given by

\begin{equation} \label{eq:lambda_0}
\lambda_0 = \frac{p_{10}(1-p_{13})}{p_{13}+p_{10}(1-p_{13})} \lambda_1 + \frac{p_{20}(1-p_{23})}{p_{23}+p_{20}(1-p_{23})} \lambda_2.
\end{equation}

The expressions for the average service rates for $S_1$, $S_2$ and $R$ are given by (\ref{eq:mu_1}), (\ref{eq:mu_2}) and (\ref{eq:mu_0}) respectively.

\begin{equation} \label{eq:mu_1}
\mu_1 = q_1 (1-q_2\mathrm{Pr}[Q_2 > 0]) (1-q_0\mathrm{Pr}[Q_0 > 0]) [p_{13}+p_{10}(1-p_{13})],
\end{equation}

\begin{equation} \label{eq:mu_2}
\mu_2 = q_2 (1-q_1\mathrm{Pr}[Q_1 > 0]) (1-q_0\mathrm{Pr}[Q_0 > 0]) [p_{23}+p_{20}(1-p_{23})],
\end{equation}

\begin{equation} \label{eq:mu_0}
\mu_0 = q_0 (1-q_1\mathrm{Pr}[Q_1 > 0]) (1-q_2\mathrm{Pr}[Q_2 > 0]) p_{03}.
\end{equation}

Since the average service rate of each queue depends on the queue size of the other queues, it cannot be computed directly. We bypass this difficulty by utilizing the idea of stochastic dominance~\cite{rao:stability}; that is, we first construct hypothetical dominant systems, in which one of the source nodes transmits dummy packets even when its packet queue is empty, while the other transmits according to its traffic. Since the queue sizes in the dominant system are, at all times, at least as large as those of the original system, the stability region of the dominant system inner-bounds that of the original system.

For the analysis in this section we use a "lower" average service rate for the relay; we assume that for the relay the users $S_1$ and $S_2$ transmit packets in every time slot with probability $q_1$ and $q_2$ respectively. As a result, (\ref{eq:mu_0}) becomes

\begin{equation}
\mu_0 = q_0 (1-q_1) (1-q_2) p_{03}.
\end{equation}

From Loynes' theorem, it turns out that the queue at the relay is stable if and only if $\lambda_0 < \mu_0$, thus

\begin{align} \label{eq:relay_stable}
\frac{p_{10}(1-p_{13})}{p_{13}+p_{10}(1-p_{13})} \lambda_1 + \frac{p_{20}(1-p_{23})}{p_{23}+p_{20}(1-p_{23})} \lambda_2 \\ < q_0 (1-q_1) (1-q_2) p_{03},
\end{align}

which is $R_{IN}^R$ in Theorem~\ref{thm:th1}.

The probability that the relay queue is non-empty is given by
\begin{equation} \label{eq:pr_relay_empty}
\mathrm{Pr}[Q_0 > 0] = \frac{\frac{p_{10}(1-p_{13})}{p_{13}+p_{10}(1-p_{13})} \lambda_1 + \frac{p_{20}(1-p_{23})}{p_{23}+p_{20}(1-p_{23})} \lambda_2}{q_0 (1-q_1) (1-q_2) p_{03}}.
\end{equation}

Note that the way we calculate the arrival rate $\lambda_0$ at the relay does not count the dummy packets, and thus we have the exact rate of the packets that arrive at the relay. The arrival rate at the relay (which is a linear combination of the arrival rates of the users) is restricted by the assumed ``lower" service rate (which is less than the actual). This is the reason why the stability region computed in this section is an inner bound.

\subsection{The first dominant system: $S_1$ user transmits dummy packets}

We obtain here the region $\mathcal{R}_{IN}^1$ of Theorem~\ref{thm:th1}.
We consider the first dominant system, in which node $S_1$ transmits dummy packets with probability $q_1$ whenever its queue is empty, while nodes $S_2$ and $R$ behave in the same way as in the original system. All other assumptions remain unaltered in the dominant system. Thus the service rate for the user $S_2$ is given by

\begin{equation} \label{eq:mu_2_dom1}
\mu_2 = q_2 (1-q_1) (1-q_0\mathrm{Pr}[Q_0 > 0]) [p_{23}+p_{20}(1-p_{23})],
\end{equation}

where $\mathrm{Pr}[Q_0 > 0]$ is given by (\ref{eq:pr_relay_empty}).

The queue at the second user is stable if and only if $\lambda_2 < \mu_2$; thus, after substituting (\ref{eq:pr_relay_empty}) into (\ref{eq:mu_2_dom1}) the stability condition is given by

\begin{equation} \label{eq:stable2_dom1}
\begin{aligned}
\frac{p_{10}(1-p_{13})}{p_{13}+p_{10}(1-p_{13})} \lambda_1 + \frac{(1-q_2)p_{03}+q_2p_{20}(1-p_{23})}{q_2[p_{23}+p_{20}(1-p_{23})]} \lambda_2  \\ < (1-q_1) (1-q_2) p_{03}.
\end{aligned}
\end{equation}

From Little's theorem we obtain that

{\footnotesize
\begin{equation} \label{eq:pr_s2_empty_dom1}
\mathrm{Pr}[Q_2>0] = \frac{\lambda_2}{q_2 (1-q_1) (1-q_0\mathrm{Pr}[Q_0 > 0]) [p_{23}+p_{20}(1-p_{23})]}.
\end{equation}
}
The queue at the first user is stable if and only if $\lambda_1 < \mu_1$, and after some algebra we obtain that

\begin{equation} \label {eq:stable1_dom1}
\begin{aligned}
\frac{(1-q_1)(1-q_2)p_{03}+q_1 p_{10}(1-p_{13})}{q_1(1-q_1)(1-q_2)p_{03}[p_{13}+p_{10}(1-p_{13})]} \lambda_1 +  \\ +\frac{(1-q_2)p_{03}+p_{20}(1-p_{23})}{(1-q_1)(1-q_2)p_{03}[p_{23}+p_{20}(1-p_{23})]} \lambda_2 < 1.
\end{aligned}
\end{equation}

The stability region obtained by the first dominant system is given by the conditions (\ref{eq:stable1_dom1}), (\ref{eq:stable2_dom1}) and (\ref{eq:relay_stable}), and is denoted by $R_{IN}^1 \bigcap R_{IN}^R$ in Theorem~\ref{thm:th1}.

\subsection{The second dominant system: $S_2$ user transmits dummy packets}
In this subsection, we obtain the region $\mathcal{R}_{IN}^2$ of Theorem~\ref{thm:th1}.
We consider the second dominant system, in which node $S_2$ transmits dummy packets with probability $q_2$ whenever its queue is empty, while nodes $S_1$ and $R$ behave in the same way as in the original system. All other assumptions remain unaltered in the dominant system.

The service rate for the $S_1$ node is given by

\begin{equation} \label{eq:mu_1_dom2}
\mu_1 = q_1 (1-q_2) (1-q_0\mathrm{Pr}[Q_0 > 0]) [p_{13}+p_{10}(1-p_{13})],
\end{equation}
where $\mathrm{Pr}[Q_0 > 0]$ is given by (\ref{eq:pr_relay_empty}). The queue at the first user is stable if and only if $\lambda_1 < \mu_1$; thus, after substituting (\ref{eq:pr_relay_empty}) into (\ref{eq:mu_1_dom2}) the stability condition is given by

\begin{equation} \label{eq:stable1_dom2}
\begin{aligned}
\frac{q_1 p_{10} (1-p_{13})+(1-q_1) p_{03}}{q_1 [p_{13}+p_{10}(1-p_{13})]} \lambda_1 + \frac{p_{20}(1-p_{23})}{p_{23}+p_{20}(1-p_{23})} \lambda_2 \\ < (1-q_1) (1-q_2) p_{03}.
\end{aligned}
\end{equation}

From Little's theorem we obtain that
{\footnotesize
\begin{equation} \label{eq:pr_s1_empty_dom2}
\mathrm{Pr}[Q_1>0] = \frac{\lambda_1}{q_1 (1-q_2) (1-q_0\mathrm{Pr}[Q_0 > 0]) [p_{13}+p_{10}(1-p_{13})]}.
\end{equation}
}
The queue at the second user is stable if and only if $\lambda_2 < \mu_2$, and after some algebra we obtain that

\begin{equation} \label {eq:stable2_dom2}
\begin{aligned}
\frac{(1-q_1)p_{03}+p_{10}(1-p_{13})}{(1-q_1)(1-q_2)p_{03}[p_{13}+p_{10}(1-p_{13})]} \lambda_1 + \\ \frac{(1-q_1)(1-q_2)p_{03}+q_2 p_{20}(1-p_{23})}{q_2 (1-q_1)(1-q_2)p_{03}[p_{23}+p_{20}(1-p_{23})]} \lambda_2 < 1.
\end{aligned}
\end{equation}

The stability region obtained by the second dominant system is given by the conditions (\ref{eq:stable2_dom2}), (\ref{eq:stable1_dom2}) and (\ref{eq:relay_stable}), and is denoted by $R_{IN}^2 \bigcap R_{IN}^R$ in Theorem~\ref{thm:th1}.

The inner stability region $R_{IN}$ (Theorem~\ref{thm:th1}) is the union of the sub-regions obtained by the two dominant systems:
\begin{align*}
R_{IN} = \left(R_{IN}^1 \bigcap R_{IN}^R \right) \bigcup \left(R_{IN}^2 \bigcap R_{IN}^R \right)= \\ = \left(R_{IN}^1 \bigcup R_{IN}^2 \right) \bigcap R_{IN}^R.
\end{align*}

\emph{Inner bound discussion:} In the first dominant system, the source $S_1$ keeps transmitting dummy packets when its queue is empty, and the actual service rate for the relay is
$\mu_0 = q_0 (1-q_1) (1-q_2\mathrm{Pr}[Q_2 > 0]) p_{03}$, instead of the assumed ``lower" $\mu_0 = q_0 (1-q_1) (1-q_2) p_{03}$. It is obvious that when the queue at the $S_2$ is almost fully utilized then the ``assumed" service rate tends to the actual one, however when $S_2$ is of low utilization then the assumed rate is smaller than the actual, this case leads to the inner bound. The same argument holds for the second dominant system.

\section{Outer Bound for the Stability Region} \label{sec:analysis_outer}

In this section, we present an outer bound for the stability region denoted by $R_{OUT}$ in Theorem~\ref{thm:th2}.
For that, we make the assumption that the relay packet transmissions do not collide with those of the sources (we can say that the users communicate in the same channel, but the relay transmits in an orthogonal to the users channel). The relay transmits a packet whenever is backlogged. The destination can receive at most two packets at the same timeslot (one from relay and one from one user). All the other assumptions remain unaltered. We provide this outer bound as an outer limit of the studied network topology.
With the previous assumptions, the average service rates for the users $S_1$, $S_2$ and the relay $R$ are given by $\mu_1 = q_1 (1-q_2\mathrm{Pr}[Q_2 > 0]) [p_{13}+p_{10}(1-p_{13})]$, $\mu_2 = q_2 (1-q_1\mathrm{Pr}[Q_1 > 0]) [p_{23}+p_{20}(1-p_{23})]$ and $\mu_0 = p_{03}$, respectively.

The average arrival rate at the relay is given by (\ref{eq:lambda_0}).
The queue at the relay is stable if $\lambda_0 < \mu_0$ as described from (\ref{eq:R_OUT_R}) in Theorem~\ref{thm:th2} .

Using the stochastic dominance argument as in Section~\ref{sec:analysis_inner}, we construct two dominant systems.
In the first dominant system, $S_1$ transmits dummy packets with probability $q_1$ when its queue is empty, whereas the others assumptions remain unaltered.
From the first system we obtain (following exactly the same steps as in Section~\ref{sec:analysis_inner}) the sub-region $R_{OUT}^{1} \bigcap R_{OUT}^R$,
where $R_{OUT}^1$ is given by (\ref{eq:R_OUT_1}).

From the second dominant system ($S_2$ transmits dummy packets), we obtain the subregion $R_{OUT}^{2} \bigcap R_{OUT}^R$.

The outer stability region $R_{OUT}$ (Theorem~\ref{thm:th2}) is the union of the sub-regions obtained by the two dominant systems, i.e.
\begin{align*}
R_{OUT} = \left(R_{OUT}^1 \bigcap R_{OUT}^R \right) \bigcup \left(R_{OUT}^2 \bigcap R_{OUT}^R \right)  =\\ = \left(R_{OUT}^1 \bigcup R_{OUT}^2 \right) \bigcap R_{OUT}^R.
\end{align*}

\section{Numerical Results} \label{sec:results}

In this section, we evaluate the stability region presenting the inner and the outer bound given in Section~\ref{sec:theorems}. We also illustrate the stability region of the network without the relay\footnote{The exact stability region for the network without the relay is known~\cite{rao:stability} and is given by $\mathcal{R}_{NOR}=\mathcal{R}^{1}_{NOR} \cup \mathcal{R}^{2}_{NOR}$, where $\mathcal{R}^{1}_{NOR}= \left\lbrace  (\lambda_{1},\lambda_{2}): \lambda_1 + \frac{q_1 p_{13}}{(1-q_1)p_{23}} \lambda_2 < q_1 p_{13}, \lambda_2 < q_2 (1-q_1)p_{23} \right\rbrace$ and $\mathcal{R}^{2}_{NOR}= \left\lbrace  (\lambda_{1},\lambda_{2}): \lambda_2 + \frac{q_2 p_{23}}{(1-q_2)p_{13}} \lambda_1 < q_2 p_{23}, \lambda_1 < q_1 (1-q_2)p_{13} \right\rbrace$.} for comparison reasons.

\begin{figure}[t]
\centering
\includegraphics[scale=0.5]{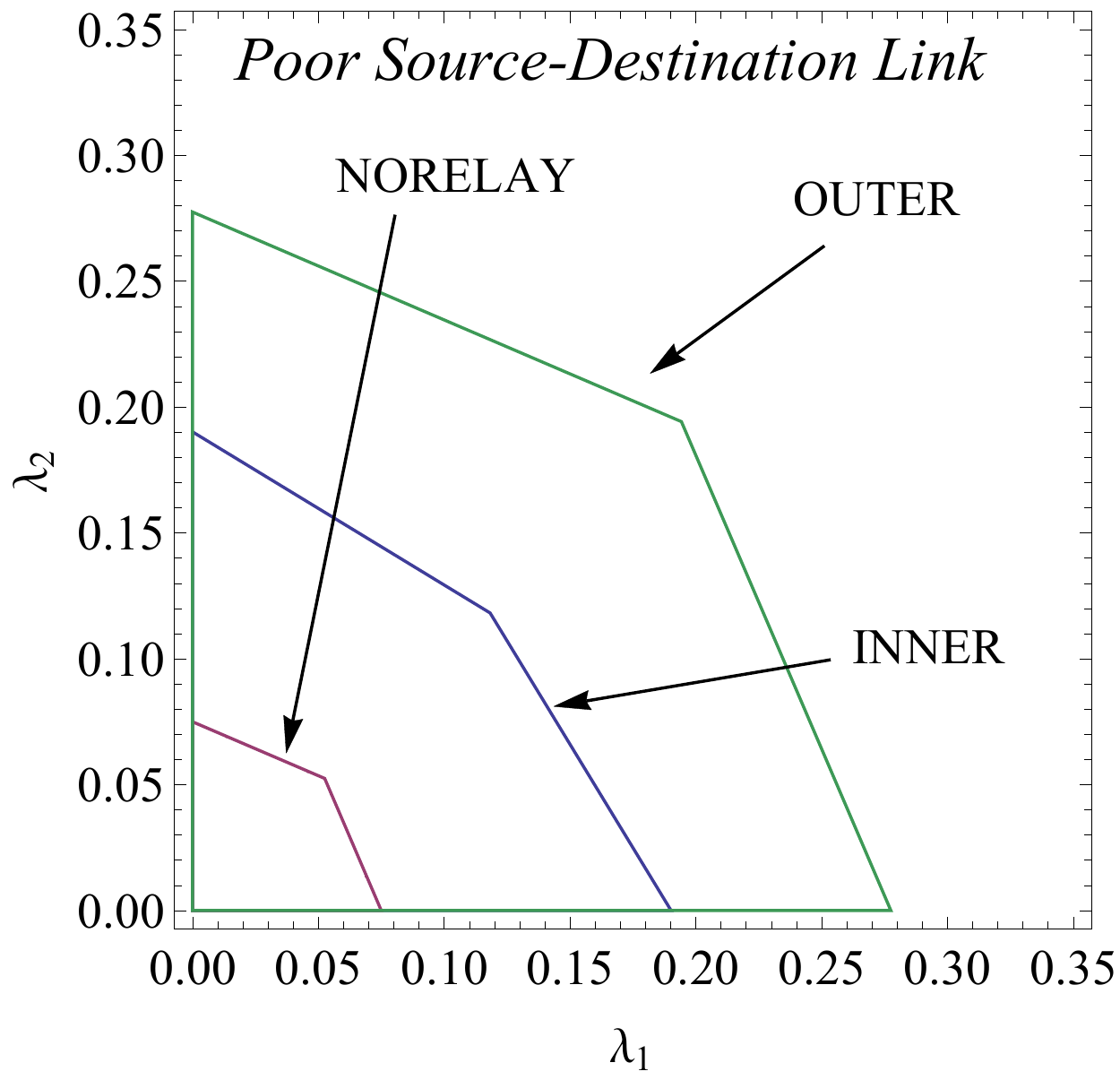}
\caption{Stability Region: $p_{13}=p_{23}=0.25$, $p_{10}=p_{20}=0.9$ and $p_{03}=0.9$.}
\label{fig:plot1}
\end{figure}

In Fig.~\ref{fig:plot1}, we depict the case where the link between the source and the destination is poor. The channel success probabilities are chosen to be $p_{13}=p_{23}=0.25$, $p_{10}=p_{20}=0.9$ and $p_{03}=0.9$. The transmission probabilities are $q_1 = q_2 = 0.3$ and $q_0 = 0.45$. The advantages offered by the existence of the relay from a stability point of view are evident. This comes in contrast with the naive thought that the existence of a relay in the collision channel could cause more collisions, which as we show is not true. Moreover, we observe that even the inner bound of the stability region offers significant gains as compared to the no-relay case.

\begin{figure}[t]
\centering
\includegraphics[scale=0.5]{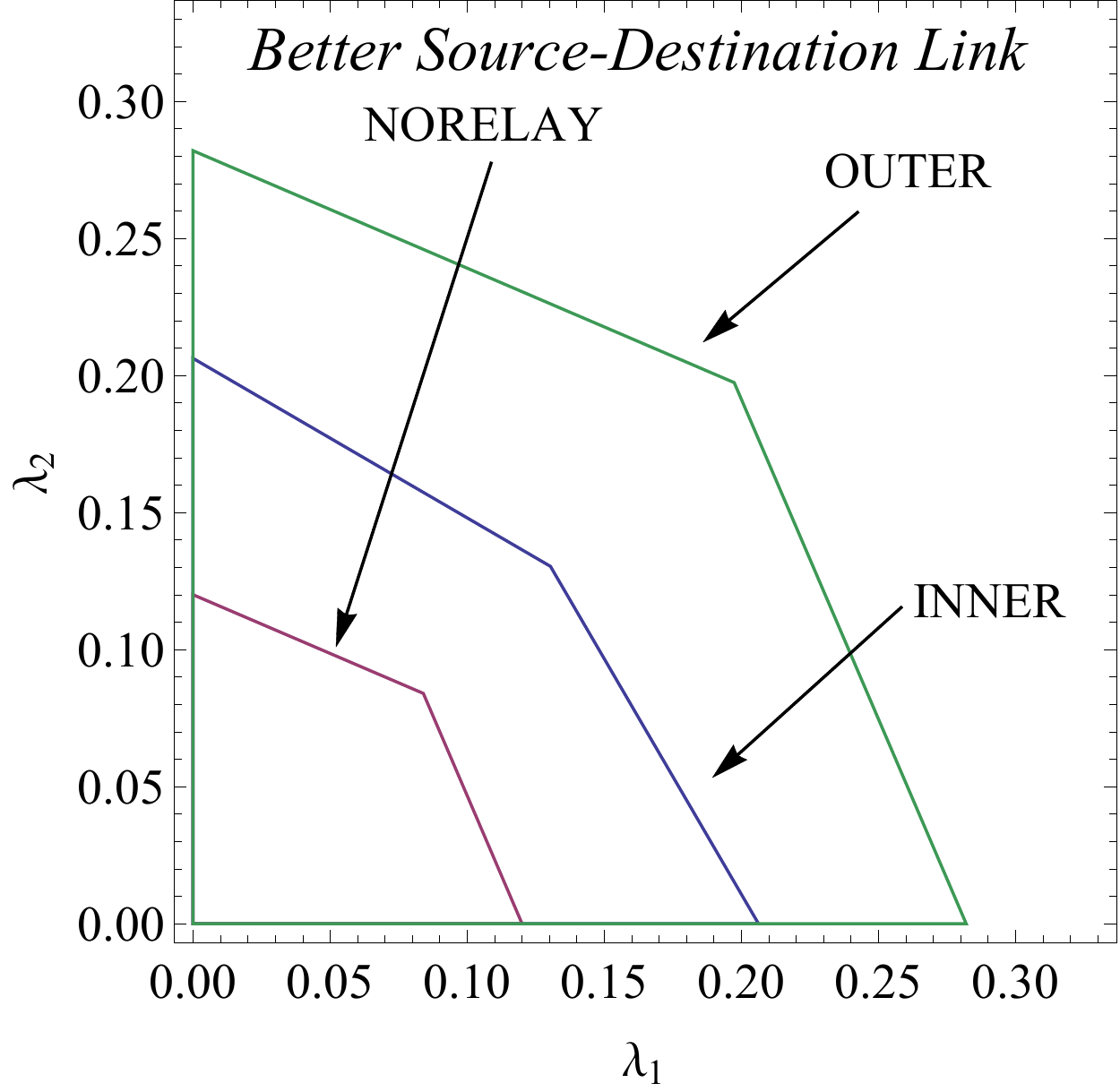}
\caption{Stability Region: $p_{13}=p_{23}=0.4$, $p_{10}=p_{20}=0.9$ and $p_{03}=0.9$.}
\label{fig:plot2}
\end{figure}

In Fig.~\ref{fig:plot2}, we study the case where the link between the source and the destination is better than the previous case. In particular, the channel success probabilities are chosen to be $p_{13}=p_{23}=0.4$, $p_{10}=p_{20}=0.9$ and $p_{03}=0.9$, and the transmission probabilities are $q_1 = q_2 = 0.3$ and $q_0 = 0.45$.

In the better-source destination link the inner bound and the stability region of the network without the relay become closer. Expectedly, as the link characteristics between the source and the destination improve then the need for the relay is of diminishing importance.

\section{Conclusions} \label{sec:conclusions}
In this work, we studied the stability region of a network which consists of two user sources with external arrivals and one relay with no packets on its own.
In particular, we obtained an inner and an outer bound for the characterization of the stability region.

The results illustrated that the deployment of the relay offers significant gains even in a collision channel model with erasures as compared to the network without the relay.

Further extensions to this work could include the generalization of the bounds to the case of $N$ users and the extension of the analysis under multi-packet reception channel model.

\bibliographystyle{IEEEtran}
\bibliography{thesis}
\end{document}